\documentclass[12pt]{article}

\usepackage{newtxtext,newtxmath}
\usepackage{graphicx,amsmath}
\usepackage{hyperref}
\usepackage{color}
\usepackage{amssymb}
\usepackage{caption}
\usepackage[a4paper,left=2.0cm,right=2.0cm,top=2.0cm,bottom=2.0cm]{geometry}
\usepackage{cite} 
\makeatletter
\def\@maketitle{%
  \newpage\setlength{\parskip}{12pt}%
    {\Large\bfseries\noindent\sloppy \@title \par}%
    {\noindent\sloppy \@author}%
}

\renewcommand\@biblabel[1]{#1.}
\def\@cite#1#2{$^{\mbox{\scriptsize #1\if@tempswa , #2\fi}}$}

\newcommand{\NAT@ignore}[2][]{}

\makeatother

\title{Night-time measurements of astronomical seeing at Dome A in Antarctica}

\author{
Bin Ma$^{1,2*}$, Zhaohui Shang$^{1,3*}$, Yi Hu$^{1*}$, Keliang Hu$^1$, 
Yongjiang Wang$^1$, Xu Yang$^1$, Michael~C.~B. Ashley$^{4}$, 
Paul Hickson$^{2}$ \& Peng Jiang$^{5}$\\
 $^1$National Astronomical Observatories, Chinese Academy of Sciences, 
Beijing 100101, China.
 $^2$Department of Physics and Astronomy, University of British Columbia, 
 6224 Agricultural Road, Vancouver, BC V6T 1Z1, Canada.
 $^3$Tianjin Astrophysics Center, Tianjin Normal University, 
Tianjin 300387, China.
 $^4$School of Physics, University of New South Wales, NSW 2052, Australia.
 $^5$Key Laboratory for Polar Science, MNR, Polar Research Institute of China, 451 Jinqiao Rd, Shanghai 200136, China.
 $^*$email: bma@nao.cas.cn; zshang@gmail.com; huyi.naoc@gmail.com\\
}

\date{}

\newcommand{\pasp}{\textit{Pub. Astron. Soc. Pacif.}}
\newcommand{\procspie}{\textit{Proc. SPIE}}

\newcommand{\aaps}{\textit{Astron. Astrophys. Suppl.}}
\newcommand{\aap}{\textit{Astron. Astrophys.}}
\newcommand{\aj}{\textit{Astron. J.}}
\newcommand{\mnras}{\textit{Mon. Not. R. Astron. Soc.}}
\newcommand{\nat}{\textit{Nature}}

\newcommand{\mps}{\,$\rm{m~s^{-1}}$}

\begin{document}

\maketitle
\vbox{}
 
\noindent {\bf 
Seeing, the angular size of stellar images blurred by atmospheric turbulence, is a critical parameter used to assess the quality of astronomical sites.  Median values at the best mid-latitude sites are generally in the range of 0.6--0.8\,arcsec\cite{2009PASP..121.1151S, 2012PASP..124..868V, 2011arXiv1101.2340T}. Sites on the Antarctic plateau are characterized by comparatively-weak turbulence in the free-atmosphere above a strong but thin boundary layer\cite{1999A&AS..134..161M, 2006PASP..118.1190S, 2009PASP..121..976S}. The median seeing at Dome C is estimated to be 0.23--0.36 arcsec\cite{2004Natur.431..278L, 2006PASP..118..344A, 2008PASP..120..203T, 2009A&A...499..955A} above a boundary layer that has a typical height of 30\,m\cite{2009A&A...499..955A, 2014A&A...568A..44P, 2015MNRAS.454.4304A}. 
At Dome A and F, the only previous seeing measurements 
were made during daytime\cite{2013IAUS..288..296P, 2013A&A...554L...5O}. 
Here we report the first direct measurements of night-time seeing at Dome A, using a Differential Image Motion Monitor\cite{1990A&A...227..294S}.  Located at a height of just 8\,m, it recorded seeing as low as 0.13\,arcsec, and provided seeing statistics that are comparable to those for a 20\,m height at Dome C. 
It indicates that the boundary layer was below 8\,m 31\% of the time. At such times the median seeing was 0.31\,arcsec, consistent with free-atmosphere seeing. 
The seeing and boundary layer thickness are found to be strongly correlated with the near-surface temperature gradient. 
The correlation confirms a median thickness of approximately 14\,m for the boundary layer at Dome A, as found from a sonic radar\cite{2010PASP..122.1122B}. 
The thinner boundary layer makes it less challenging to locate a telescope above it, thereby giving greater access to the free-atmosphere.
}

To directly measure the seeing at Dome A, the location of Kunlun Station (see Extended Data Fig.\,1), we developed the KunLun 
Differential Image Motion Monitors (KL-DIMMs\cite{2018SPIE10700E..52M}). They were designed to operate automatically at temperatures as low as $-80^\circ$C. 
In January 2019, the 35th Chinese National Antarctic Research Expedition installed 
two KL-DIMMs on an 8\,m tower. 
The KL-DIMM provided a seeing measurement every minute. In total, we obtained 45,930 valid seeing values during the polar night from 11 April to 4 August 2019. These values were corrected to the zenith and a wavelength of 500\,nm, as is standard. 
During observations, frost and/or snow accumulated on the two wedges that form the sub-apertures. The frost reduced not only the signal but also  
the effective diameters of the sub-apertures. 
We decide to use a fixed diameter value of 5\,cm for all seeing estimates. Thus, the derived seeing values are upper limits; the true values  
are likely 10--20\% lower (see Methods).

\begin{figure}
\centering
\includegraphics[width=12cm]{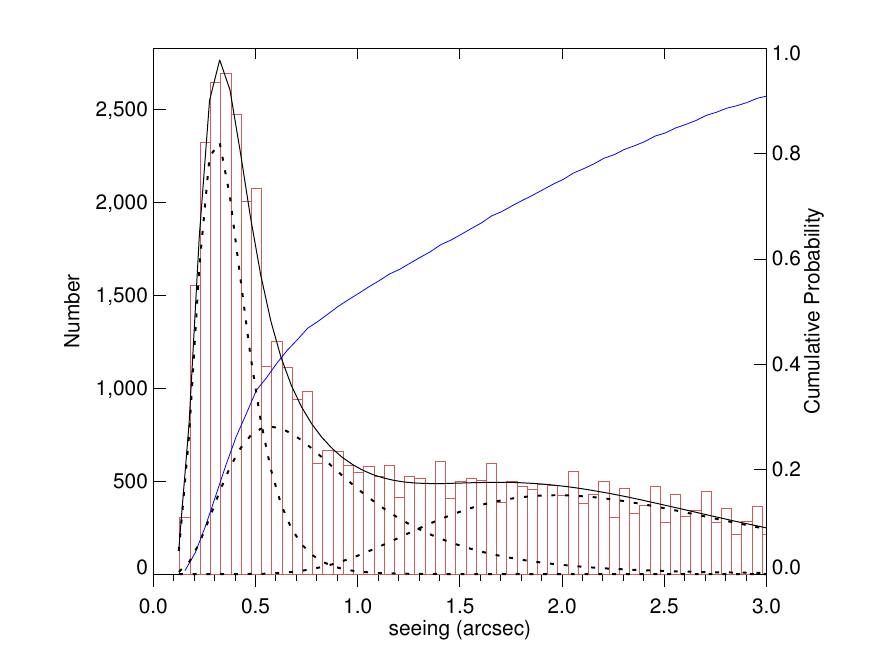}
\caption{
\textbf{$|$ The night-time seeing at Dome A in 2019.} 
The histogram is in brown and the cumulative probability is in blue. 
The solid black line fits the histogram with a sum of three log-normal  
components (dotted black lines) representing the FA seeing, the intermediate seeing and the 
BL seeing from small to large values (see Table \ref{tab:seeing}). 
\label{fig:hist}}
\end{figure}

Fig.\,\ref{fig:hist} shows the histogram of seeing from Dome A during the polar night in 
2019. The distribution consists of a strikingly-sharp peak that has a mode value of 0.31 arcsec, as well as a long tail of poor seeing extending beyond 3 arcsec. 
The peak at 0.31 arcsec is on order of the expected free-atmosphere (FA) seeing, thus it is reasonable to attribute the times of superb and poor seeing to the cases when the boundary layer (BL) was entirely below or extended above the telescope, respectively. In addition, we expect an intermediate case\cite{2009A&A...499..955A} when there is a transition phase of the BL thickness varying across the apertures of the KL-DIMM, or a secondary turbulent layer above the surface layer. 
We therefore fit the histogram with three log-normal components\cite{2009A&A...499..955A}, corresponding to the 
FA, intermediate and BL components, as shown in Fig.\,\ref{fig:hist}.
The fraction of FA seeing is consistent with the results from the sonic radar\cite{2010PASP..122.1122B}.

We compare the statistics of seeing from Dome A with those from Dome C\cite{2009A&A...499..955A} and mid-latitude sites in 
Table \ref{tab:seeing}. For the height of 8\,m, the seeing statistics at Dome A are much better than those at the same height from Dome C, and are comparable to those at 20\,m from Dome C. This is because the fraction of FA seeing is higher at Dome A, although the FA seeing values are similar.  
The comparison confirms that the BL at Dome A is much thinner, and more turbulent, than at Dome C, as was predicted by models\cite{2006PASP..118.1190S}.
We therefore expect that on a tower that is higher than 8\,m at Dome A, superb seeing will be more frequent while poor seeing will be reduced in both frequency and value, as is the case at Dome C. 
At the best mid-latitude sites, the BL contributes less to the seeing, but 
it has a thickness that can be hundreds of metres and therefore cannot be avoided. In Antarctica, 
a telescope on an adequate tower would be affected by the FA only, enabling 
high-resolution wide-field observations.

To estimate the seeing at other heights, we can correlate the seeing, or BL thickness, with near-ground meteorological parameters, since most turbulence resides within a few tens-of-metres of the snow surface.
In 2019 the expedition 
installed a new second-generation KunLun Automated Weather Station\cite{2019PASP..131a5001H}, which measured temperature, wind speed and wind direction every 
2\,m from the snow surface to a height of 14\,m, as well as air pressure and relative humidity. 
However, we only have 1.5 months of simultaneous data with KL-DIMM, mainly during the polar day, up until 15 March.

A comparison of seeing and weather data (see Methods)
reveals that the best seeing occurs when the temperature gradient is large. 
In a temperature inversion, warm air overlies cooler air resulting in a stable atmosphere. However, wind shear may generate turbulence. The relative importance of buoyancy and wind shear is quantified by a dimensionless ratio, $Ri$, the Richardson number. If $Ri$ is below the critical value of $0.25$, turbulence develops.
Even for 0.25 $< Ri <$ 1, residual turbulence may remain for some time during the transition from turbulent to laminar flow\cite{2002ASPC..266....2V}. 

\begin{figure}
\centering
\includegraphics[width=8.0cm]{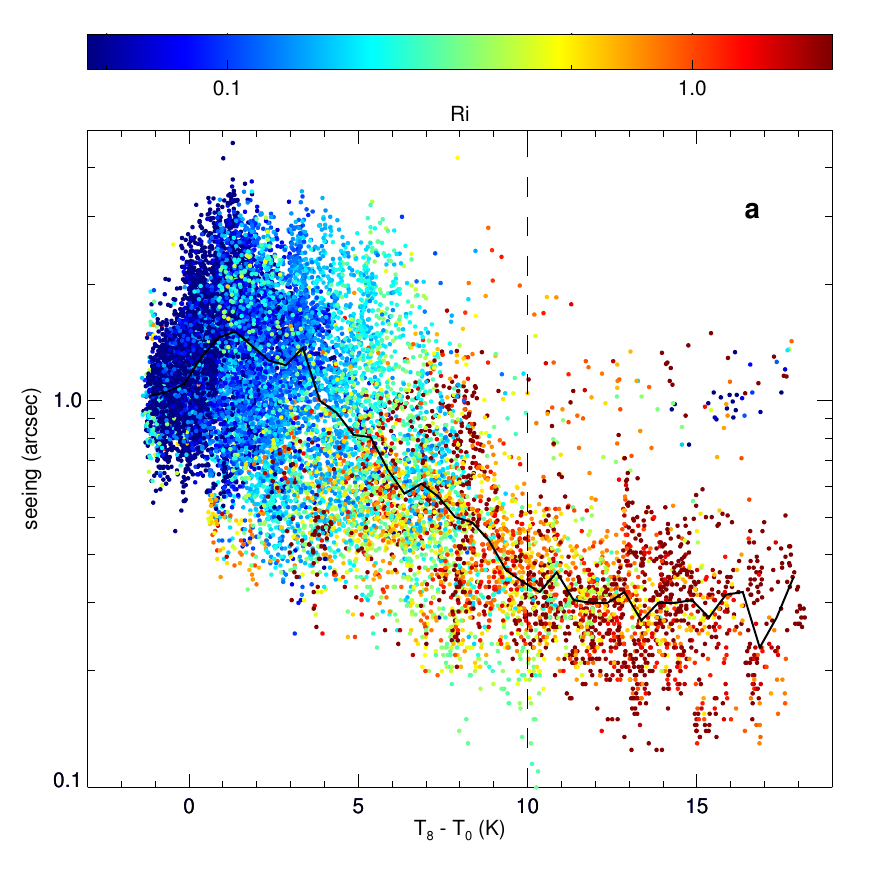} 
\includegraphics[width=8.0cm]{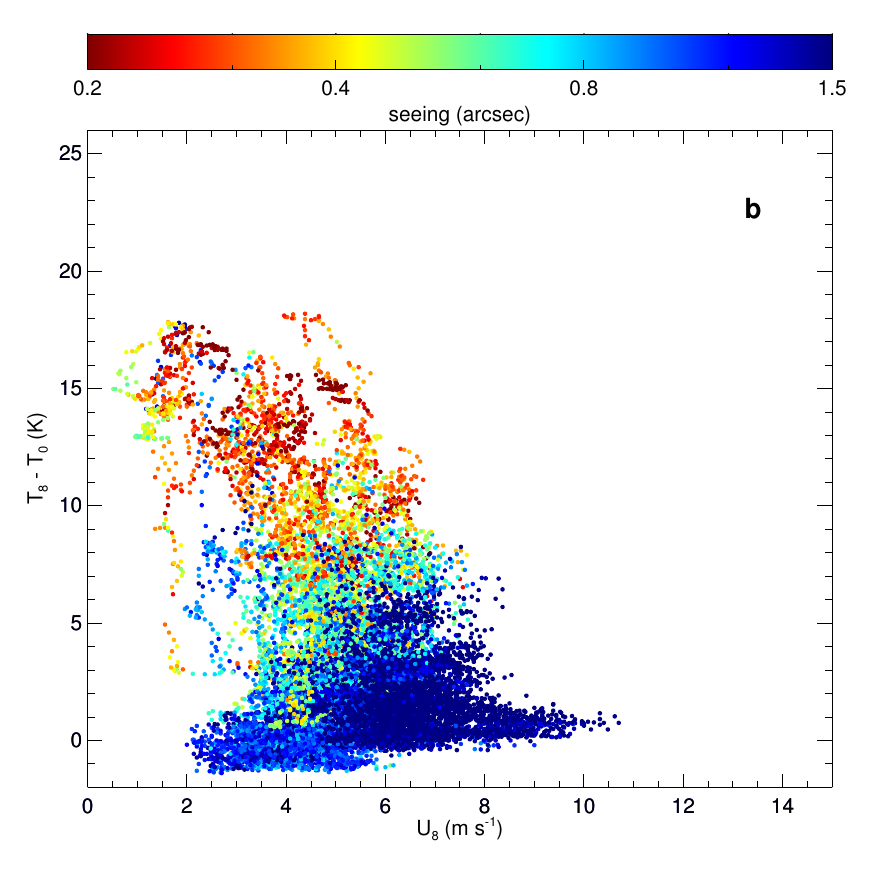} 
\caption{\textbf{$|$ The correlations between seeing and meteorological parameters.} {\bf a,} The strong correlation between seeing and $T_8 - T_0$. Colour indicates $Ri$. The solid line is the median seeing for each 0.5\,K bin of $T_8 - T_0$, and the dashed line indicates $T_8 - T_0 = 10$\,K. 
{\bf b,} $T_8 - T_0$ versus wind speed at 8\,m, $U_8$, with colour indicating seeing.
\label{fig:seeing-weather}}
\end{figure}

In Fig.\,\ref{fig:seeing-weather}a, it can be seen that the seeing at the height of 8\,m depends strongly on the temperature difference between 8\,m and 0\,m (surface), $T_8 - T_0$, which dominates $Ri$ at 8\,m. Three regions of different behaviour can be identified:
\begin{enumerate}
\setlength{\itemsep}{-1ex}
\item $T_8 - T_0 < 1$\,K, $Ri \sim$ 0, the median seeing increases from 1 to 1.5\,arcsec;
\item as $T_8 - T_0$ increases from 1 to 10\,K, the median $Ri$ increases from 0 to 0.8, and the median seeing decreases from 1.5 to 0.3\,arcsec;
\item $T_8 - T_0 > 10$\,K, $Ri > 0.8$, seeing remains constant at $\sim$ 0.3\,arcsec for most data but with a small fraction having seeing $\sim$ 1\,arcsec. 
\end{enumerate}
Because seeing close to 0.3 arcsec suggests that the KL-DIMM is  above the BL, 
the dependence of seeing on $T_8 - T_0$ reflects the height variation of the  BL. Similar dependence at Dome C is also reported\cite{2019BoLMe.171..101P}. 
As $T_8 - T_0$ increases, so does $Ri$. The turbulence intensity then decreases with height rapidly and the BL becomes thinner. 
Consequently, the seeing at 8\,m improves until the BL is below 8\,m, 
when the seeing is determined only by the FA.  The seeing has little dependence on $T_8 - T_0$ beyond 10\,K, which is the critical value when the thickness of the BL is $\sim$ 8\,m. 
Occasionally, when extra high-altitude turbulence occurs, the seeing is degraded to $\sim 1$\,arcsec, which is also observed at Dome C\cite{2019BoLMe.171..101P}.

Although the seeing is correlated mainly with $T_8 - T_0$, the scatter is quite large ($\sim$ 0.2\,dex). From Fig.\,\ref{fig:seeing-weather}b, we find that 
the wind speed at 8\,m, $U_8$, also plays a role. There is a clear trend of improved seeing with decreasing $U_8$ for small $T_8-T_0$. Good seeing and large $T_8-T_0$ never occur when $U_8 \gtrsim 7$ \mps. 
Modelling also indicates that the surface wind speed is an indicator for 
BL thickness and seeing\cite{2006PASP..118.1190S}.

\begin{figure*}
\centering
\includegraphics[width=8.0cm]{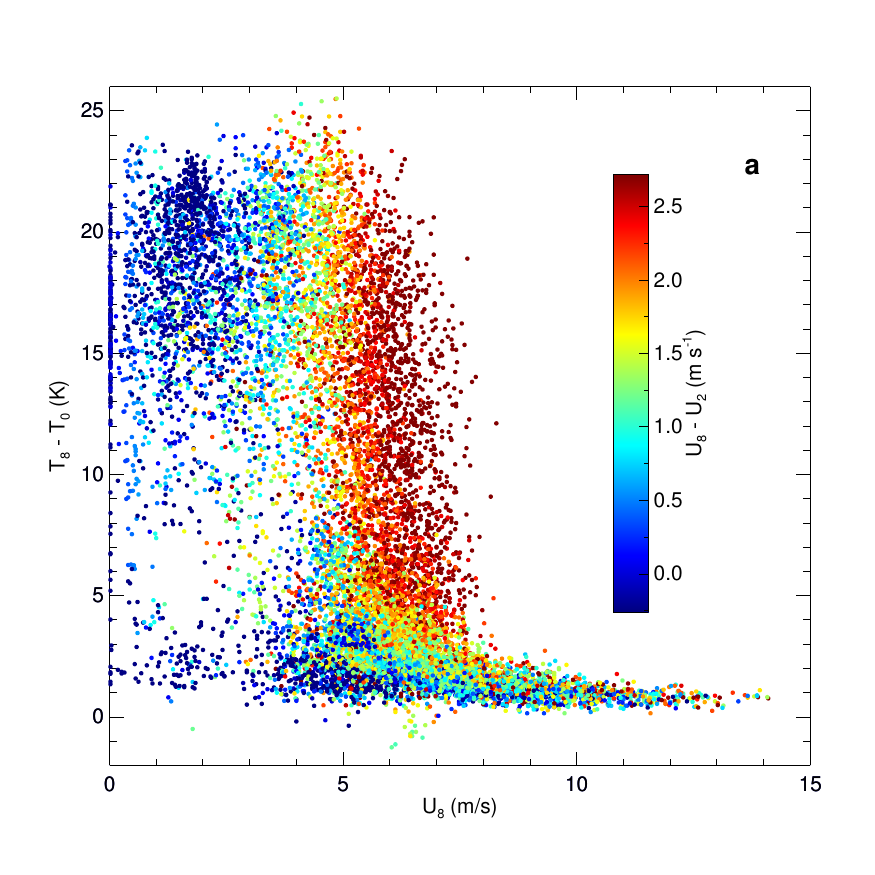}
\includegraphics[width=8.0cm]{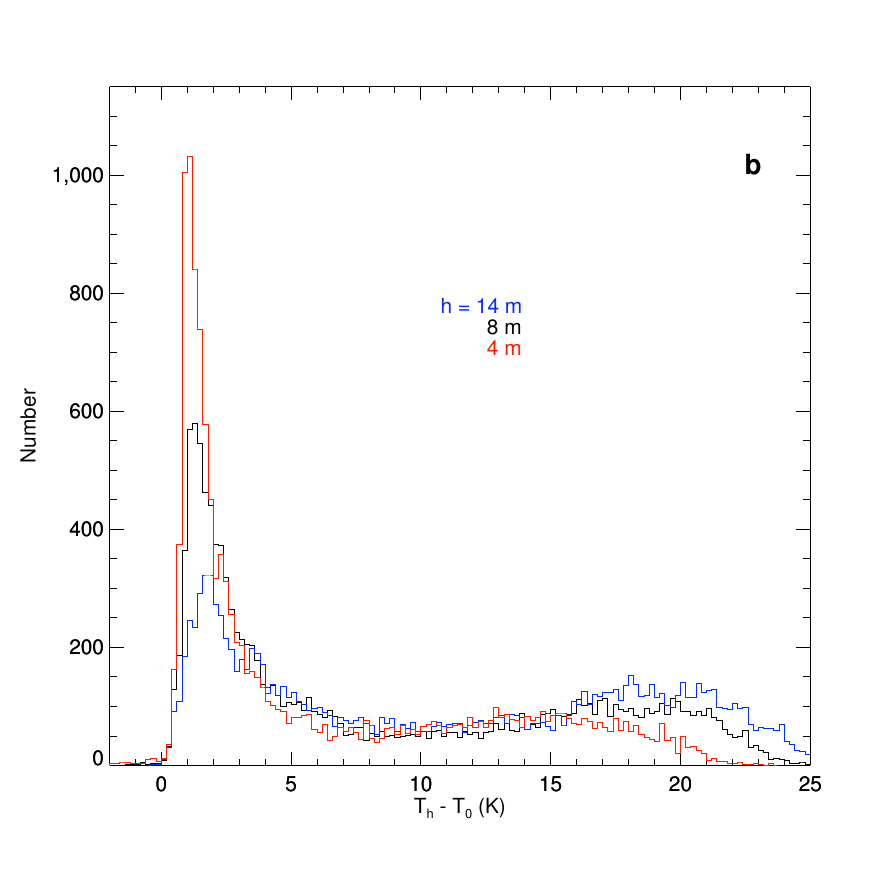}
\caption{\textbf{$|$ The statistics of polar-night weather data in 2015.} {\bf a,} $T_8 - T_0$ 
against $U_8$ as Fig.\,\ref{fig:seeing-weather}{\bf b}, with the colours indicating 
the wind speed difference between 8\,m and 2\,m, $U_8 - U_2$. 
The data points concentrate in two horizontal regimes: large $T_8 - T_0$ with $U_8 \lesssim$  7\mps, and small $T_8 - T_0$ with $U_8$ mostly from 3 to 15\mps. Also, they are connected by a vertical zone where $U_8 \sim$ 6\mps. 
{\bf b,} The histograms of temperature differences between 
$T_h$ at the height of $h$ and $T_0$ at the surface, for $h = 14, 8$ and 4\,m, 
respectively. 
\label{fig:weather2015}}
\end{figure*}

To infer the dependence of the BL on meteorological parameters during the polar night, we analyzed historical weather data\cite{2019PASP..131a5001H}. Fig.\,\ref{fig:weather2015}a shows the same plot as Fig.\,\ref{fig:seeing-weather}b, but for the winter period from April to August in 2015. 
The two regimes with large and small $T_8 - T_0$ correspond to the thin and thick BL cases, respectively. 
Also, the fraction of large $T_8 - T_0$ increases significantly compared to the polar daytime data.
Strikingly, the greatest $U_8 - U_2$ does not occur with the greatest $U_8$, but occurs during the transitional phase with moderate $U_8$. 
The meteorological data from Dome C show a similar pattern, and two regimes separated by a $U_{10}$ threshold have been proposed\cite{2017QJRMS.143.1241V}. However, our results 
prefer the temperature difference as an indicator to distinguish the two regimes (or BL thickness). 

In principle, the temperature difference, $T_h - T_0$, between the height $h$ and the surface could be used to estimate the fraction of time with good seeing for $h$.
Then, one could optimize the height for the installation of a large telescope, balancing the seeing quality and the difficulty of construction. 
We plot the histograms of $T_h - T_0$ for $h = 4, 8 \text{ and } 14$\,m, respectively, in Fig.\,\ref{fig:weather2015}b for the polar night in 2015.
The distributions are bimodal, corresponding to the BL below and above $h$. 
$T_8 - T_0 > $ 10\,K, suggesting that the FA seeing would be obtained at an 8 m height, occurred 38\% of the time. This agrees with the KL-DIMM result for the polar night in 2019. 
For $h = 14$\,m, the two peaks of the histogram are larger than those for 8\,m by less than 1\,K. Thus we approximately set the threshold for $T_{14} - T_0$ to be 11\,K, 1\,K greater than that for $T_8 - T_0$. The fraction of time with $T_{14} - T_0 >$ 11\,K, when the BL thickness is expected to be lower than 14\,m, was 49\% in 2015 . This is also consistent with previous measurements\cite{2010PASP..122.1122B}.

In conclusion, the exceptional seeing, in combination with clear and dark sky\cite{2010AJ....140..602Z, 2017AJ....154....6Y}, and the low thermal-infrared background\cite{1996PASP..108..721A} from the coldest air\cite{2014PASP..126..868H, 2019PASP..131a5001H}, would significantly improve both angular resolutions and limiting magnitudes of optical/infrared telescopes at Dome A. 
The proposed 2.5\,m Kunlun Dark Universe Survey Telescope (KDUST\cite{2013IAUS..288..271Y}) would be able to compete with 6\,m class telescopes at the best mid-latitude sites.  
In addition, Dome A is a natural laboratory for studies of the formation and dissipation of 
turbulence within the boundary layer. Future measurements of weather, seeing, and the 
low-altitude turbulence profile\cite{2019MNRAS.485.2532H}, could contribute to a better understanding of the Antarctic 
atmosphere\cite{2015ACP....15.6225G, 2017JGRD..122.6818V, 2019QJRMS.145..930B}.



\begin{table*}[!p]
\begin{center}
\caption{\bf$|$ Statistics of the night-time seeing}
\label{tab:seeing}
\resizebox{\textwidth}{!}{ 
\begin{tabular}{lccccccccccc}
\hline
Site  & height & 25\% &  median  & 75\% & $\epsilon_{FA}$ & $\epsilon_{IN}$ & $\epsilon_{BL}$ & $f_{FA}$ & $f_{IN}$ & $f_{BL}$\\ 
 & m & arcsec  & arcsec & arcsec & arcsec & arcsec & arcsec & \% & \% & \% \\
\hline
Dome A & 8 &  0.41  &  0.89  &  2.02 & 0.31 & 0.57 & 1.97 & 31.0 & 30.1 & 38.9\\ 
Dome C & 8 &  0.83  &  1.65  &  2.32 & 0.33 & 0.54 & 1.73 & 16.2 & 14.4 & 69.4\\
Dome C & 20 &  0.43 & 0.84   &  1.55 & 0.30 & 0.42 & 1.17 & 15.7 & 29.3 & 55.0\\
Mauna Kea & 7 & 0.57 & 0.75 & 1.03 \\
Armazones & 7 & 0.50 & 0.64 & 0.86 \\
La Palma & 5 & 0.62 & 0.80 & 1.06 \\
\hline
\end{tabular}}
\end{center}
The columns are the names of sites, the heights of towers for DIMMs, the 25th, 50th (median) and 75th percentile values of DIMM seeing, the median values of the FA seeing $\epsilon_{FA}$, intermediate seeing $\epsilon_{IN}$ and BL seeing $\epsilon_{BL}$, and the fractions of time with the FA seeing $f_{FA}$, intermediate seeing $f_{IN}$ and BL seeing $f_{BL}$, respectively. 
The entrance pupils of DIMMs are usually $\sim$ 1\,m above the top of towers. The last six columns are the decomposition results for three log-normal components in seeing histograms and only for Dome A/C. 
For Dome C\cite{2009A&A...499..955A}, the seeing values at 8\,m are from winter observations. 
We also list the seeing at best mid-latitude sites: Mauna Kea 13N in Hawaii\cite{2009PASP..121.1151S}, Cerro Armazones in Chile\cite{2009PASP..121.1151S} and La Palma in Canary Islands\cite{2012PASP..124..868V}.

\end{table*}

\clearpage
\newpage

\clearpage

\section*{Methods}
\subsection*{KL-DIMM}

We developed two KL-DIMMs for Dome A, which were almost identical, for redundancy. Both KL-DIMMs passed a cold test to $-70^\circ$C and functioned properly. They were also cross-calibrated by a well-established DIMM in a western China site, Muztagh-ata\cite{2020arXiv200313998X}. And the seeing measurements were consistent, with a  standard deviation of 10\% for the relative differences. 

At Dome A, two KL-DIMMs were supported by PLATO-A\cite{2010SPIE.7735E..40A}, a logistical system that provides power and internet access. 
 Although one of them produced little data owing to tracking and focusing problems, the other one functioned well even when the air temperature was below $-70^\circ$C. 
The latter obtained data automatically from 25 January to 11 March, and from 11 April to 4 August.

For deicing/defrosting during the polar night, power was supplied to the indium-tin-oxide (ITO) conductive coatings on the wedges, but residual snow or frost still could be seen in web-camera images. The frost reduced
the effective diameters, $D_{\text{eff}}$, of the sub-apertures, which is a parameter in the seeing calculation.
A possible approach to correcting for this effect is to assume 
$D_{\text{eff}} = \sqrt{f_{\text{obs}} / {f_0}} D_0$, 
where $f_{\text{obs}}$ is the observed stellar flux, $f_0$ is the flux with no frost, and $D_0 = 5$\,cm is the diameter of the wedges. The extinction from 
clouds was generally very small according to the simultaneous images from an all-sky camera\cite{2018SPIE10700E..57S}, thus most extinction was expected to be caused by frost and snow on the apertures.
The median ratios of $f_{\text{obs}} / {f_0}$ were 0.27 and 0.47 for the two sub-apertures, respectively. 
Smaller $D_{\text{eff}}$ results in greater image motion\cite{1990A&A...227..294S, 2002PASP..114.1156T}, causing the seeing to be overestimated. For a flux ratio of 0.27 (0.47) the seeing is overestimated by 24 (14) per cent. 
However, the shapes of the real sub-apertures are irregular and unknown, so this effect cannot be corrected precisely. We therefore
  used a fixed value of 5\,cm for all seeing estimates.

\subsection*{Seeing-weather correlation}
An example of seeing and weather data from 4 to 5 March 2019 is presented in Extended Data Fig.\,2. The seeing shows a strong diurnal variation, ranging from $\sim 0.3$\,arcsec at local midnight and $\sim 1$\,arcsec at noon. Sporadic bursts of seeing variations are also apparent on timescales as short as 1 hour.  
On these days, the air temperature at the snow surface, $T_0$, also varied diurnally. It was about $-55^\circ$C at night, rising quickly to about $-45^\circ$C after sunrise, then falling again to $-55^\circ$C after sunset. 
 Smaller variations are seen in the temperatures $T_8$ and $T_{12}$, at 8\,m and 12\,m heights, which were typically close to $-40^\circ$C and showed little diurnal trend. 
These two heights are adopted because 8\,m is the height of the KL-DIMM and 12\,m is the highest that has a valid temperature measurement (the sensor at 14\,m was faulty). 
A strong temperature inversion developed rapidly at night, with the temperature difference between 8\,m and the snow surface ($T_8 - T_0$) often exceeding 15$^\circ$C. The temperature gradient generally decreased rapidly with height. The inversion decreased or even disappeared during the day, due to solar heating.
Yet, the temperature inversion was not always strong at night. 
For instance, on the night of 5 March when $T_0$ decreased rapidly, $T_8$ and $T_{12}$ were also decreasing, due to the strong wind and wind speed gradient. At that time, the seeing was not improved from day to night. 
Besides the diurnal trend, the rapid variations of seeing were also accompanied by rapid  temperature inversion variations, possibly due to the fast variation of wind direction. 
This example implies that the seeing is strongly correlated with the intensity of temperature inversion. 

\subsection*{Data availability}
The seeing and weather data at Dome A in 2019 that support the findings of this study
are available in China-VO Paper Data Repository, http://paperdata.china-vo.org/BinMa/DomeA-seeing2019.zip. 
The weather data in 2015 have been published\cite{2019PASP..131a5001H} and are available in http://aag.bao.ac.cn/klaws/downloads/.
The data are also displayed on a public website, http://aag.bao.ac.cn/klsite/klaws2g.php.



\setlength{\parindent}{0em} 

{\bf Acknowledgements} The authors are grateful to the 35th Chinese National 
 Antarctic Research Expedition team supported by the Polar Research Institute of China and the Chinese Arctic and Antarctic Administration. 
 We also thank Drs. Zhengyang Li, Fujia Du, Zhongwen Hu, Tengfei Song, Eric Aristidi, Karim Agabi for their help to the development of the KL-DIMMs. 
 We thank the Chinese Antarctic Center of Surveying and Mapping for providing the map of Antarctica.
 This work is supported by the National Natural Science Foundation of China under grant numbers 11733007, 11673037 and 11873010,
 and is partly supported by  the  Operation,  Maintenance  and  Upgrading  Fund  for Astronomical Telescopes  and  Facility  Instruments,  budgeted  from  the  Ministry  of Finance of China (MOF) and administrated by the Chinese Academy of Sciences (CAS).
  The tower for KL-DIMMs was funded by Tianjin Normal University, and implemented by Tianjin University.
  PH acknowledges support from the Natural Sciences and Engineering Research Council of Canada, RGPIN-2019-04369, and the Canada Foundation for Innovation. MA acknowledges support from the Australian Antarctic Division and NCRIS funding through Astronomy Australia Limited.

\vbox{} 

{\bf Author contributions}
 B.M., Z.S., Y.H., K.H., Y.W., and X.Y. contributed to the development and tests of 
 the KL-DIMM. 
 K.H., Z.S. and P.J. installed KL-DIMM at Dome A, Antarctica, while 
 B.M.,Y.H., Y.W. and X.Y. contributed to the remote operation. 
 M.A. contributed to the PLATO-A platform supplying power and internet access.
 B.M. analysed the data and wrote the manuscript together with P.H., Z.S and M.A.
 All authors reviewed and commented on the manuscript.

\vbox{}

{\bf Competing Interests} The authors declare no competing interests.

\vbox{}

 {\bf Additional Information}\\
 {\bf Correspondence and requests for materials} should be addressed to
 B.M., Z.S. or Y.H.\\
 {\bf Reprints and permissions information}is available at www.nature.com/reprints.


\clearpage
\newpage

\begin{figure}
\centering
\includegraphics[width=15.0cm]{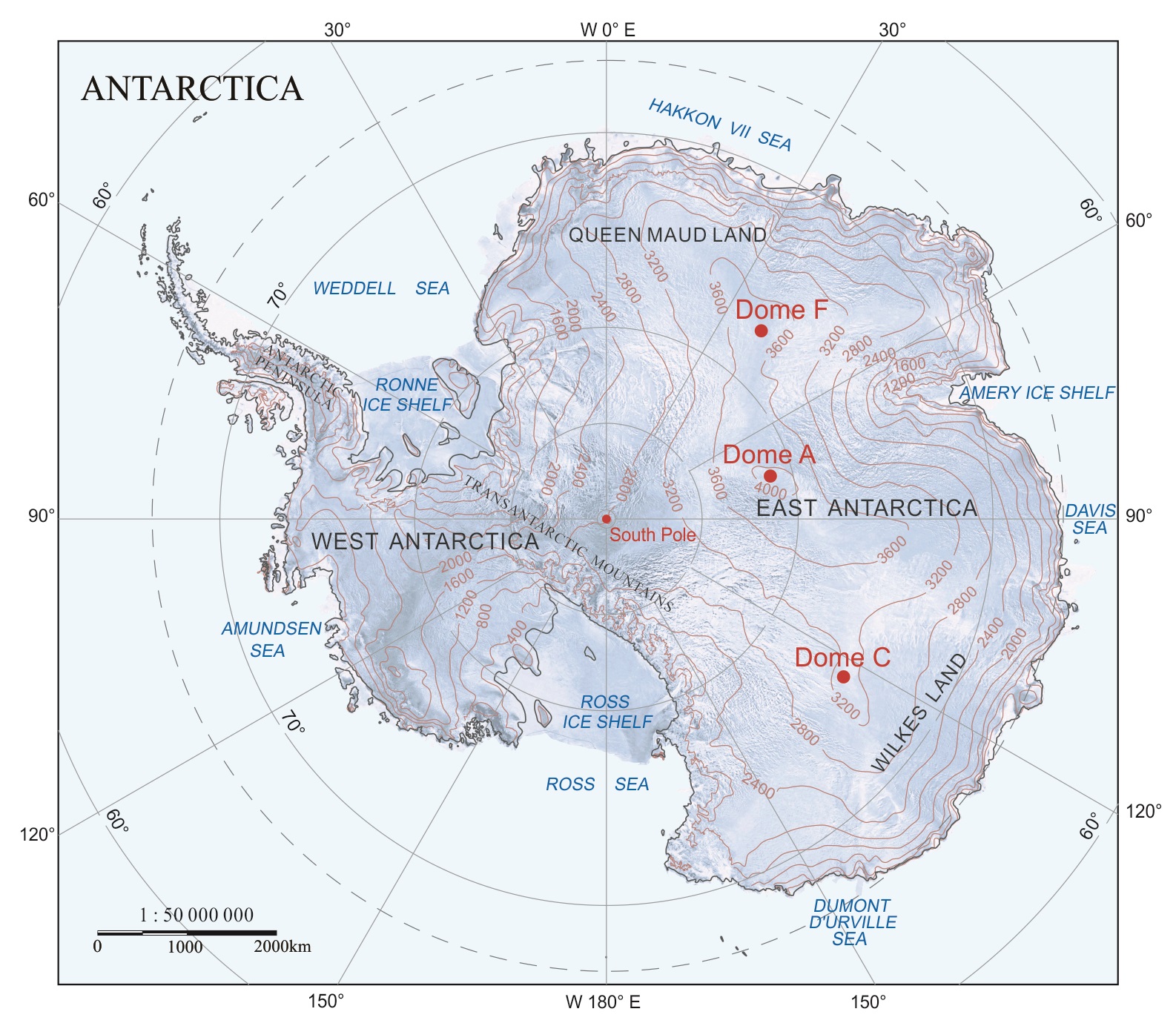}
\caption*{{\bf Extended Data Fig.\,1 $|$ Map of Antarctica.} The red dots indicate locations of Dome A (Chinese Kunlun Station), Dome C (French–Italian Concordia Station), Dome F (Japanese Dome Fuji Station) and the South Pole (Unite States Amundsen–Scott Station), respectively. Courtesy of Xiaoping Pang and Shiyun Wang. 
\label{fig:map}}
\end{figure}

\begin{figure}
\centering
\includegraphics[width=13.0cm]{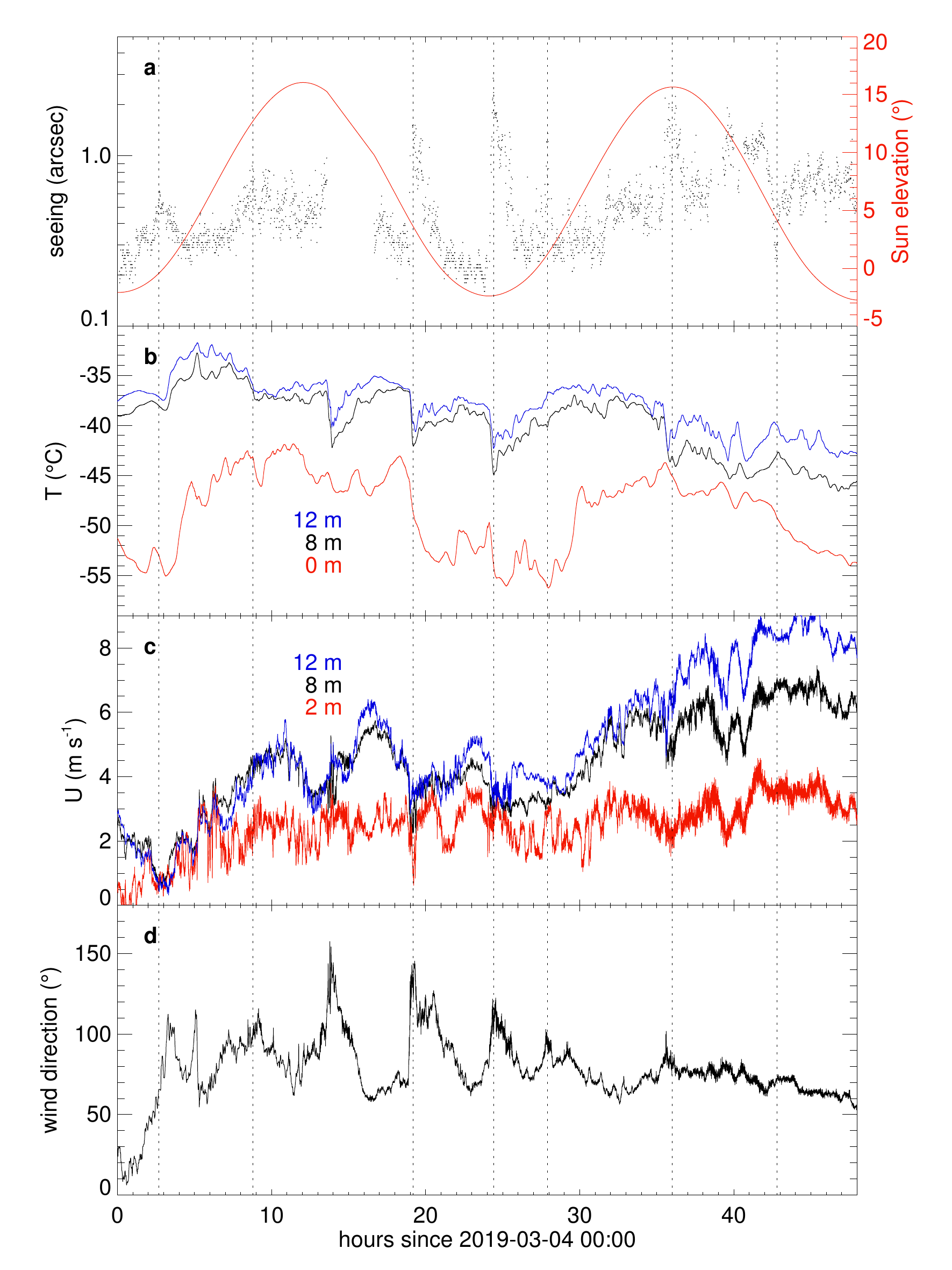}
\caption*{{\bf Extended Data Fig.\,2 $|$ An example of seeing and meteorological data from 4 and 5 March 2019.} {\bf a,} Seeing (black) and Sun elevation (red). {\bf b,} Air temperature at 12, 8 and 0\,m heights. {\bf c,} Wind speed at 12, 8 and 2\,m. {\bf d,}  Wind direction at 8\,m. Dotted vertical lines mark times when seeing either increased or decreased rapidly. 
\label{fig:seeing-weather-eg}}
\end{figure}


\begin{thebibliography}{1}
\setlength{\itemsep}{-1ex} 

\bibitem{2009PASP..121.1151S} Skidmore, W., et al. Thirty Meter Telescope Site Testing V: Seeing and Isoplanatic Angle.\ \pasp\ \textbf{121}, 1151 (2009).
\bibitem{2012PASP..124..868V} V{\'a}zquez Rami{\'o}, H., et al. European Extremely Large Telescope Site Characterization. II. High Angular Resolution Parameters.\ \pasp\ \textbf{124}, 868 (2012).
\bibitem{2011arXiv1101.2340T} Thomas-Osip, J.~E., McCarthy, P., Prieto, G., Phillips, M.~M., Johns, M. Giant Magellan Telescope Site Testing Summary.\ arXiv e-prints\  arXiv:1101.2340 (2011).


\bibitem{1999A&AS..134..161M} Marks, R.~D., Vernin, J., Azouit, M., Manigault, J.~F., Clevelin, C. Measurement of optical seeing on the high antarctic plateau.\ \aaps\ \textbf{134}, 161-172 (1999).
\bibitem{2006PASP..118.1190S} Swain, M.~R., Gall{\'e}e, H. Antarctic Boundary Layer Seeing.\ \pasp\ \textbf{118}, 1190-1197 (2006).
\bibitem{2009PASP..121..976S} Saunders, W., et al. Where Is the Best Site on Earth? Domes A, B, C, and F, and Ridges A and B.\ \pasp\ \textbf{121}, 976 (2009).

\bibitem{2004Natur.431..278L} Lawrence, J.~S., Ashley, M.~C.~B., Tokovinin, A., Travouillon, T. Exceptional astronomical seeing conditions above Dome C in Antarctica.\ \nat\ \textbf{431}, 278-281 (2004).
\bibitem{2006PASP..118..344A} Agabi, A., et al. First Whole Atmosphere night-time Seeing Measurements at Dome C, Antarctica.\ \pasp\ \textbf{118}, 344-348 (2006).
\bibitem{2008PASP..120..203T} Trinquet, H., et al. Nighttime Optical Turbulence Vertical Structure above Dome C in Antarctica.\ \pasp\ \textbf{120}, 203 (2008).
\bibitem{2009A&A...499..955A} Aristidi, E., et al. Dome C site testing: surface layer, free atmosphere seeing, and isoplanatic angle statistics.\ \aap\ \textbf{499}, 955-965 (2009).

\bibitem{2014A&A...568A..44P} Petenko, I., et al. Observations of optically active turbulence in the planetary boundary layer by sodar at the Concordia astronomical observatory, Dome C, Antarctica.\ \aap\ \textbf{568}, A44 (2014).
\bibitem{2015MNRAS.454.4304A} Aristidi, E., et al. Monitoring the optical turbulence in the surface layer at Dome C, Antarctica, with sonic anemometers.\ \mnras\ \textbf{454}, 4304-4315 (2015).

\bibitem{2013IAUS..288..296P} Pei, C., Li, Z., Chen, H., Yuan, X. Preliminary daytime seeing monitoring at Dome A, Antarctica.\ Astrophysics from Antarctica\ \textbf{288}, 296-297 (2013).
\bibitem{2013A&A...554L...5O} Okita, H., Ichikawa, T., Ashley, M.~C.~B., Takato, N., Motoyama, H. Excellent daytime seeing at Dome Fuji on the Antarctic plateau.\ \aap\ \textbf{554}, L5 (2013).
\bibitem{1990A&A...227..294S} Sarazin, M., Roddier, F. The ESO differential image motion monitor.\ \aap\ \textbf{227}, 294-300 (1990).
\bibitem{2010PASP..122.1122B} Bonner, C.~S., et al. Thickness of the Atmospheric Boundary Layer Above Dome A, Antarctica, during 2009.\ \pasp\ \textbf{122}, 1122 (2010).
\bibitem{2018SPIE10700E..52M} Ma, B., et al. An automatic DIMM for Dome A, Antarctica.\ \procspie\ \textbf{10700}, 1070052 (2018).

\bibitem{2019PASP..131a5001H} Hu, Y., et al. Meteorological Data from KLAWS-2G for an Astronomical Site Survey of Dome A, Antarctica.\ \pasp\ \textbf{131}, 015001 (2019).
\bibitem{2002ASPC..266....2V} Vernin, J. Mechanism of formation of optical turbulence (Invited Speaker).\ Astronomical Site Evaluation in the Visible and Radio Range\ \textbf{266}, 2 (2002).

\bibitem{2019BoLMe.171..101P} Petenko, I., Argentini, S., Casasanta, G., Genthon, C., Kallistratova, M. Stable Surface-Based Turbulent Layer During the Polar Winter at Dome C, Antarctica: Sodar and In Situ Observations.\ Boundary-Layer Meteorology\ \textbf{171}, 101-128 (2019).
\bibitem{2017QJRMS.143.1241V} Vignon, E., et al. Stable boundary-layer regimes at Dome C, Antarctica: observation and analysis.\ Quarterly Journal of the Royal Meteorological Society\ \textbf{143}, 1241-1253 (2017).

\bibitem{2010AJ....140..602Z} Zou, H., et al. Sky Brightness and Transparency in the i-band at Dome A, Antarctica.\ \aj\ \textbf{140}, 602-611 (2010).
\bibitem{2017AJ....154....6Y} Yang, Y., et al. Optical Sky Brightness and Transparency during the Winter Season at Dome A Antarctica from the Gattini-All-Sky Camera.\ \aj\ \textbf{154}, 6 (2017).
\bibitem{1996PASP..108..721A} Ashley, M.~C.~B., et al. South Pole Observations of the Near-Infrared Sky Brightness.\ \pasp\ \textbf{108}, 721 (1996).

\bibitem{2014PASP..126..868H} Hu, Y., et al. Meteorological Data for the Astronomical Site at Dome A, Antarctica.\ \pasp\ \textbf{126}, 868 (2014).

\bibitem{2013IAUS..288..271Y} Yuan, X., et al. Preliminary design of the Kunlun Dark Universe Survey Telescope (KDUST).\ Astrophysics from Antarctica\ \textbf{288}, 271-274 (2013).
\bibitem{2019MNRAS.485.2532H} Hickson, P., Ma, B., Shang, Z., Xue, S. Multistar turbulence monitor: a new technique to measure optical turbulence profiles.\ \mnras\ \textbf{485}, 2532-2545 (2019).
\bibitem{2015ACP....15.6225G} Gall{\'e}e, H., et al. Characterization of the boundary layer at Dome C (East Antarctica) during the OPALE summer campaign.\ Atmospheric Chemistry \& Physics\ \textbf{15}, 6225-6236 (2015).
\bibitem{2017JGRD..122.6818V} Vignon, E., et al. Antarctic boundary layer parametrization in a general circulation model: 1-D simulations facing summer observations at Dome C.\ Journal of Geophysical Research (Atmospheres)\ \textbf{122}, 6818-6843 (2017).
\bibitem{2019QJRMS.145..930B} Baas, P., et al. Transitions in the wintertime near-surface temperature inversion at Dome C, Antarctica.\ Quarterly Journal of the Royal Meteorological Society\ \textbf{145}, 930-946 (2019).

\end{thebibliography}

\begin{thebibliography}{1}
\setlength{\itemsep}{-1ex} 

\bibitem[31]{2020arXiv200313998X} Xu, J., et al. Site-testing at Muztagh-ata site II: Seeing statistics.\ arXiv e-prints\  arXiv:2003.13998 (2020).

\bibitem[32]{2010SPIE.7735E..40A} Ashley, M.~C.~B., et al. Future development of the PLATO Observatory for Antarctic science.\ \procspie\ \textbf{7735}, 773540 (2010).
\bibitem[33]{2018SPIE10700E..57S} Shang, Z., et al. Kunlun cloud and aurora monitor.\ \procspie\ \textbf{10700}, 1070057 (2018).
\bibitem[34]{2002PASP..114.1156T} Tokovinin, A. From Differential Image Motion to Seeing.\ \pasp\ \textbf{114}, 1156-1166 (2002).

\end{thebibliography}
\end{document}